\documentclass[conference, 10pt]{IEEEtran}
\IEEEoverridecommandlockouts
\usepackage{latexsym}
\usepackage{graphicx}
\usepackage{amsfonts,amssymb,amsmath}

\def\nbc{{\mathbf{c}}}

\def\nbu{{\mathbf{u}}}
\def\nbv{{\mathbf{v}}}

\def\nbx{{\mathbf{x}}}

\def\nbz{{\mathbf{z}}}
\def\nb0{{\mathbf{0}}}
\def\nb1{{\mathbf{1}}}

\def\nbC{{\mathbf{C}}}

\def\nbG{{\mathbf{G}}}

\def\N{\sigma^2}

\usepackage[acronyms,nonumberlist,nopostdot,nomain,nogroupskip]{glossaries}
\newacronym{quic}{QUIC}{Quick UDP Internet Connections}
\newacronym{3gpp}{3GPP}{3rd Generation Partnership Project}
\newacronym{adc}{ADC}{Analog to Digital Converter}
\newacronym{5g}{5G}{5th generation}
\newacronym{aimd}{AIMD}{Additive Increase Multiplicative Decrease}
\newacronym{am}{AM}{Acknowledged Mode}
\newacronym{amc}{AMC}{Adaptive Modulation and Coding}
\newacronym{aqm}{AQM}{Active Queue Management}
\newacronym{awgn}{AGWN}{Additive White Gaussian Noise}
\newacronym{afd}{AFD}{Austin Fire Department}
\newacronym{balia}{BALIA}{Balanced Link Adaptation}
\newacronym{bdp}{BDP}{Bandwidth-Delay Product}
\newacronym{bf}{BF}{Beamforming}
\newacronym{cc}{CC}{Congestion Control}
\newacronym{cdf}{CDF}{Cumulative Distribution Function}
\newacronym{cn}{CN}{Core Network}
\newacronym{cqi}{CQI}{Channel Quality Information}
\newacronym{cp}{CP}{Control Plane}
\newacronym{csirs}{CSI-RS}{Channel State Information - Reference Signal}
\newacronym{dc}{DC}{Dual Connectivity}
\newacronym{dce}{DCE}{Direct Code Execution}
\newacronym{dci}{DCI}{Downlink Control Information}
\newacronym{dl}{DL}{Downlink}
\newacronym{dmr}{DMR}{Deadline Miss Ratio}
\newacronym{dmrs}{DMRS}{DeModulation Reference Signal}
\newacronym{e2e}{E2E}{End-to-End}
\newacronym{ecn}{ECN}{Explicit Congestion Notification}
\newacronym{edf}{EDF}{Earliest Deadline First}
\newacronym{enb}{eNB}{evolved Node Base}
\newacronym{epc}{EPC}{Evolved Packet Core}
\newacronym{es}{ES}{Edge Server}
\newacronym{fdma}{FDMA}{Frequency Division Multiple Access}
\newacronym{fdd}{FDD}{Frequency Division Duplexing}
\newacronym[firstplural=Radio Access Technologies (RATs)]{rat}{RAT}{Radio Access Technology}
\newacronym{fs}{FS}{Fast Switching}
\newacronym{ftp}{FTP}{File Transfer Protocol}
\newacronym{gnb}{gNB}{Next Generation Node Base}
\newacronym{harq}{HARQ}{Hybrid Automatic Repeat reQuest}
\newacronym{hetnet}{HetNet}{Heterogeneous Network}
\newacronym{hh}{HH}{Hard Handover}
\newacronym{hol}{HOL}{Head-of-Line}
\newacronym{ia}{IA}{Initial Access}
\newacronym{imt}{IMT}{International Mobile Telecommunication}
\newacronym{iot}{IoT}{Internet of Things}
\newacronym{los}{LOS}{Line of Sight}
\newacronym{lte}{LTE}{Long Term Evolution}
\newacronym{m2m}{M2M}{Machine to Machine}
\newacronym{mac}{MAC}{Medium Access Control}
\newacronym{mc}{MC}{Multi-Connectivity}
\newacronym{mcs}{MCS}{Modulation and Coding Scheme}
\newacronym{mec}{MEC}{Mobile Edge Cloud}
\newacronym{mi}{MI}{Mutual Information}
\newacronym{mimo}{MIMO}{Multiple Input, Multiple Output}
\newacronym{mmwave}{mmWave}{millimeter wave}
\newacronym{mr}{MR}{Maximum Rate}
\newacronym{mss}{MSS}{Maximum Segment Size}
\newacronym{mtd}{MTD}{Machine-Type Device}
\newacronym{mtu}{MTU}{Maximum Transmission Unit}
\newacronym{nfv}{NFV}{Network Function Virtualization}
\newacronym{nlos}{NLOS}{Non Line of Sight}
\newacronym{nr}{NR}{New Radio}
\newacronym{ofdm}{OFDM}{Orthogonal Frequency Division Multiplexing}
\newacronym{pdcch}{PDCCH}{Physical Downlonk Control Channel}
\newacronym{pdcp}{PDCP}{Packet Data Convergence Protocol}
\newacronym{pdsch}{PDSCH}{Physical Downlink Shared Channel}
\newacronym{pdu}{PDU}{Packet Data Unit}
\newacronym{pf}{PF}{Proportional Fair}
\newacronym{pgw}{PGW}{Packet Gateway}
\newacronym{phy}{PHY}{Physical}
\newacronym{pbch}{PBCH}{Physical Broadcast Channel}
\newacronym[plural=\gls{mme}s,firstplural=Mobility Management Entities (MMEs)]{mme}{MME}{Mobility Management Entity}
\newacronym{prb}{PRB}{Physical Resource Block}
\newacronym{pss}{PSS}{Primary Synchronization Signal}
\newacronym{pucch}{PUCCH}{Physical Uplink Control Channel}
\newacronym{pusch}{PUSCH}{Physical Uplink Shared Channel}
\newacronym{rach}{RACH}{Random Access Channel}
\newacronym{ran}{RAN}{Radio Access Network}
\newacronym{red}{RED}{Robotics Emergency Deployment}
\newacronym{rf}{RF}{Radio Frequency}
\newacronym{rlc}{RLC}{Radio Link Control}
\newacronym{rlf}{RLF}{Radio Link Failure}
\newacronym{rrc}{RRC}{Radio Resource Control}
\newacronym{rrm}{RRM}{Radio Resource Management}
\newacronym{rr}{RR}{Round Robin}
\newacronym{rs}{RS}{Remote Server}
\newacronym{rsrp}{RSRP}{Reference Signal Received Power}
\newacronym{rss}{RSS}{Received Signal Strength}
\newacronym{rtt}{RTT}{Round Trip Time}
\newacronym{rw}{RW}{Receive Window}
\newacronym{rx}{RX}{Receiver}
\newacronym{sa}{SA}{standalone}
\newacronym{sack}{SACK}{Selective Acknowledgment}
\newacronym{sap}{SAP}{Service Access Point}
\newacronym{sch}{SCH}{Secondary Cell Handover}
\newacronym{scoot}{SCOOT}{Split Cycle Offset Optimization Technique}
\newacronym{sdma}{SDMA}{Spatial Division Multiple Access}
\newacronym{sinr}{SINR}{Signal to Interference plus Noise Ratio}
\newacronym{sm}{SM}{Saturation Mode}
\newacronym{snr}{SNR}{Signal to Noise Ratio}
\newacronym{son}{SON}{Self-Organizing Network}
\newacronym{ss}{SS}{Synchronization Signal}
\newacronym{srs}{SRS}{Sounding Reference Signal}
\newacronym{sss}{SSS}{Secondary Synchronization Signal}
\newacronym{tb}{TB}{Transport Block}
\newacronym{tcp}{TCP}{Transmission Control Protocol}
\newacronym{tdd}{TDD}{Time Division Duplexing}
\newacronym{tdma}{TDMA}{Time Division Multiple Access}
\newacronym{tfl}{TfL}{Transport for London}
\newacronym{tm}{TM}{Transparent Mode}
\newacronym{trp}{TRP}{Transmitter Receiver Pair}
\newacronym{tti}{TTI}{Transmission Time Interval}
\newacronym{ttt}{TTT}{Time-to-Trigger}
\newacronym{tx}{TX}{Transmitter}
\newacronym{ue}{UE}{User Equipment}
\newacronym{ul}{UL}{Uplink}
\newacronym{uml}{UML}{Unified Modeling Language}
\newacronym{um}{UM}{Unacknowledged Mode}
\newacronym{utc}{UTC}{Urban Traffic Control}
\newacronym{vm}{VM}{Virtual Machine}
\newacronym{rsrq}{RSRQ}{Reference Signal Received Quality}
\newacronym{rssi}{RSSI}{Received Signal Strength Indicator}
\newacronym{crs}{CRS}{Cell Reference Signal}
\newacronym{comp}{CoMP}{Coordinated Multi-Point}
\newacronym{cran}{C-RAN}{Cloud \acrlong{ran}}
\newacronym{ca}{CA}{Carrier Aggregation}
\newacronym{cco}{CC}{Carrier Component}
\newacronym{nsa}{NSA}{Non Stand Alone}
\newacronym{embb}{eMBB}{Enhanced Mobility Broadband}
\newacronym{bsr}{BSR}{Buffer Status Report}
\newacronym{srb}{SRB}{Service Radio Bearer}
\newacronym{scm}{SCM}{Spatial Channel Model}
\newacronym{sctp}{SCTP}{Stream Control Transmission Protocol}
\newacronym{mptcp}{MPTCP}{Multi-path TCP}
\newacronym{ietf}{IETF}{Internet Engineering Task Force}
\newacronym{os}{OS}{Operating System}
\newacronym{tls}{TLS}{Transport Layer Security}
\newacronym{rfc}{RFC}{Request for Comments}
\newacronym{http}{HTTP}{HyperText Transfer Protocol}
\newacronym{nat}{NAT}{Network Address Translation}
\newacronym{api}{API}{Application Programming Interface}
\newacronym{rto}{RTO}{Retransmission Timeout}
\newacronym{psc}{PSC}{Public Safety Communication}
\newacronym{rpgm}{RPGM}{Reference Point Group Mobility}
\newacronym{ic}{IC}{Incident Command}
\newacronym{rsu}{RSU}{Road Side Unit}
\newacronym{uav}{UAV}{unmanned aerial vehicle}
\newacronym{usv}{USV}{Unmanned Surface Vehicle}
\newacronym{uas}{UAS}{Unmanned Aerial System}
\newacronym{iab}{IAB}{Integrated Access and Backhaul}
\newacronym{qoe}{QoE}{Quality of Experience}
\newacronym{ssim}{SSIM}{Structural Similarity Index}
\newacronym{psnr}{PSNR}{Peak Signal to Noise Ratio}
\newacronym{bs}{BS}{Base Station}
\newacronym{mu}{MU}{Multiple User}
\newacronym{ag}{AG}{Air-to-Ground}
\newacronym{af}{AF}{Array Factor}
\newacronym{ula}{ULA}{Uniform Linear Array}
\newacronym{upa}{UPA}{Uniform Planar Array}
\newacronym{lcs}{LCS}{Local Coordinate System}
\newacronym{psd}{PSD}{Power Spectral Density}
\newacronym{vq}{VQ}{vector quantization}
\newacronym{a2g}{A2G}{air-to-ground}
\newacronym{em}{EM}{electromagnetic}
\newacronym{vae}{VAE}{variational autoencoder}

\def\bb0{{\mathbb{0}}}

\def\bb{{\boldsymbol{b}}}

\def\b0{{\boldsymbol{0}}}

\def\b{{\mathrm{b}}}

\def\r0{{\mathbf{0}}}

\def\bsf0{{\bm{\mathsf{0}}}}

\def\N0{{N_{\mathrm{0}}}}

\def\bsf{{\boldsymbol{s}_\mathrm{f}}}

\newcommand{\be}{\begin{equation}}
\newcommand{\ee}{\end{equation}}
\newcommand{\bal}{\begin{align}}
\newcommand{\eal}{\end{align}}

\usepackage{textcomp}
\usepackage{gensymb}
\usepackage{siunitx}
\usepackage{xcolor}
\usepackage{tikz}
\usepackage{etoolbox}
\usepackage{enumitem}
\usepackage{esvect}
\usepackage{mathtools}
\newtoggle{conf}
\toggletrue{conf}
\usetikzlibrary{chains,arrows,calc,positioning}
\usepackage{multirow}
\usepackage{comment}
\usepackage[caption=false, font=footnotesize]{subfig}
\usepackage[normalem]{ulem}

\newtoggle{conference}
\toggletrue{conference}  

\def\BibTeX{{\rm B\kern-.05em{\sc i\kern-.025em b}\kern-.08em T\kern-.1667em\lower.7ex\hbox{E}\kern-.125emX}}

\begin{document}
%
\title{Multi-Frequency Channel Modeling for\\
Millimeter Wave and THz Wireless Communication\\
via  Generative Adversarial Networks}
\author{
\IEEEauthorblockN{Yaqi Hu, ~Mingsheng Yin, ~William Xia,
~Sundeep Rangan, ~Marco Mezzavilla} 
\IEEEauthorblockA{NYU Tandon School of Engineering, Brooklyn, NY, USA} 
\thanks{The authors were supported 
by NSF grants 
1952180, 1925079, 1564142, 1547332, the SRC, OPPO,
and the industrial affiliates of NYU WIRELESS. The work was also supported by Remcom that provided the Wireless Insite  
software.}
}

\maketitle

\begin{abstract}
Modern cellular systems
rely increasingly on simultaneous communication in multiple discontinuous bands
for macro-diversity and increased bandwidth.  
Multi-frequency communication is particularly crucial in the
millimeter wave (mmWave) and Terahertz (THz) frequencies,
as these bands are often coupled with lower frequencies
for robustness.
Evaluation of these systems
requires statistical models that can capture the joint distribution of the channel paths across multiple frequencies.  
This paper presents a general neural network based methodology for training multi-frequency double directional statistical channel models.  
In the proposed approach, each is described as a multi-clustered set, and a generative adversarial network (GAN) is trained to generate random multi-cluster profiles where the generated cluster data includes the angles and delay of the clusters along with the vectors of random received powers, angular, and delay spread at different frequencies.  
The model can be readily applied for multi-frequency link or network layer simulation.  
The methodology is demonstrated on modeling urban micro-cellular links at \SI{28}{} and \SI{140}{GHz} trained from extensive ray tracing data.
The methodology makes
minimal statistical assumptions and experiments show the model can capture
interesting statistical relationships between frequencies.

\end{abstract}


\IEEEpeerreviewmaketitle

\section{Introduction}
\label{sec:intro}
Cellular wireless systems often  
operate over multiple frequency bands 
simultaneously for 
increased bandwidth and macro-diversity.
Commercial cellular system support several mechanisms for
multi-band operation including
carrier aggregation 
\cite{shen2012overview}, dual connectivity
\cite{yilmaz2019overview}, and multi-connectivity
\cite{ghosh20195g}.
The simultaneous use of multiple frequencies is particularly important
in the millimeter wave (mmWave) and terahertz (THz) bands
where lower frequency carriers are needed for robustness and control 
signaling
\cite{lopez2019opportunities,liu20205g}.
There has also been active research
in leveraging information from lower frequencies 
for communication in the  mmWave and THz bands
for procedures such as beam search 
\cite{gonzalez2017millimeter,hashemi2017out,giordani2019standalone},
and localization and sensing
\cite{yang2021integrated}.  

Evaluation of these systems, necessitates channel models that can describe the behavior of links
across multiple frequency bands. 
In this work, we seek to develop
\emph{statistical channel models} 
that can generate random instances of the channel 
parameters following the observed
statistical distribution of channel parameters in some environment (e.g.\ urban micro).
Statistical models such as 
those by 3GPP \cite{3GPP38901} 
are the mainstay
for commercial cellular evaluation 
and are a critical initial step in 
any simulation evaluation.

Developing accurate statistical channel 
models for mmWave systems is well known
to be challenging, even for a single band.  Since mmWave systems
operate at very high bandwidths with highly
directional beams, statistical models
must capture the full \emph{double directional}
nature of the channel meaning the angles
of arrival and departure, delays and gains of 
all the paths \cite{heath2018foundations}.  This difficulty is compounded in 
the multi-frequency setting, where 
one must generate gains of the paths across multiple frequencies.  These parameters
can have complex relationships as factors that
influence propagation, such as transmission
and reflection losses, vary considerably 
across frequency \cite{Rappaport2014-mmwbook}.

Due to the difficulty in modeling 
wireless channels from first principles,
there has been a growing interest in
using data driven, neural networks (NN) based techniques
\cite{stocker1993neural,chang1997environment,bai2018predicting,huang2018big,9122601,XiaRanMez2020}.
The key benefit of these approaches is that
they make minimal assumptions and can thus
learn complex statistical relationships
between variables.
In this line, the work in \cite{XiaRanMez2020} 
used NNs for air-to-ground channel models
at \SI{28}{GHz}.  The approach was able to model the full double-directional nature of the channel
and demonstrated the ability 
to learn non-obvious relationships.
In addition, the trained models considerably outperformed
standard 3GPP models such as \cite{3GPP38901},
which assume very specific structures to be
tractable.

In this work, we wish to model
the \emph{joint} distribution of the paths
between a transmitter (TX) and receiver (RX)
across multiple frequencies.
The channel is modeled by a set of clusters and a generative model is then used to generate the cluster parameters from the distance between the TX and RX.
The paper then develops and evaluates 
several new features as compared to the modelling methodology of \cite{XiaRanMez2020}
to support multiple frequencies:
\begin{itemize}
\item \emph{Multi-frequency cluster models:}
Most importantly, the model represents the channel in a compact
manner that can capture the key commonalities and differences
between frequencies.  Specifically, 
the channel is represented by a set of clusters
whose mean angles of arrivals and departure
and delay are common among the frequencies, reflecting that macro-level
path of propagation are not frequency dependent.  On the other hand,
the power, and the angular and delay spread are modeled as frequency
dependent to capture the  variations in scattering
and transmission losses across the frequencies,
which is particularly vital to model in the mmWave
and THz regimes \cite{degli2007measurement,schecklman2011terahertz,ju2019scattering,xing2021millimeter}.


\item \emph{Ray tracing processing:}
There are limited number of multi-frequency
channel measurements \cite{huang2020multi,cui2020multi} 
and these generally have an insufficient
numbers of data points for training large NNs.
Similar to \cite{XiaRanMez2020,khawaja2017uav,Alkhateeb2019},
we thus employ a powerful ray tracing package, Wireless InSite by Remcom \cite{Remcom}, to obtain high volume training data.

\item \emph{Cluster identification and pre-processing:}
When using the ray tracing data, we propose a novel clustering
algorithm as a pre-processing step that identifies
common clusters across frequencies but captures the delay
and angular variations in different frequencies.  
The methods can be considered as a multi-frequency
extension of clustering techniques proposed in
\cite{he2018clustering}.


\item \emph{Evaluation on micro-cellular
urban environment}:  The method is evaluated 
on micro-cellular
terrestrial links  in a region($\approx 0.6 \times 0.4$\, \si{km}$^2$) of Beijing, China.  
It is shown that the method can capture
important correlations in the channel characteristics
between frequencies.


\end{itemize}

\section{Multi-Frequency Modeling Problem}
\label{sec:model}

We consider the modeling of channels on $M$ frequencies,
$f_1,\ldots,f_M$. For example, in the examples below,
we will look at two frequencies:  $f_1=$\, \SI{28}{GHz}
and $f_2=$\, \SI{140}{GHz}.  We model a single link,
meaning the channel from a transmitter (TX) to receiver (RX).
For the cellular applications below, the TX will be 
the base station (gNB) and the RX will be the mobile
user equipment (UE).  
However, due to reciprocity, the TX and RX can be reversed.  
We model the channel across the $M$ frequencies by a common set of $L$ clusters where each cluster $\ell$ is described by a vector of parameters,
\begin{equation} \label{eq:thetaell}
    \nbx^{(\ell)}
        = \left[\tilde{\tau}_\ell, 
        \bar{\theta}^{\rm rx}_\ell, 
        \bar{\phi}^{\rm rx}_\ell, 
        \bar{\theta}^{\rm tx}_\ell, 
        \bar{\phi}^{\rm tx}_\ell,
        F_{\ell 1},\ldots, F_{\ell M}
        \right],
\end{equation}
where $\tilde{\tau}_\ell$ is the minimum path delay within the cluster,
$\bar{\theta}^{\rm rx}_\ell$, $\bar{\phi}^{\rm rx}_\ell$ are the average intra-cluster inclination and azimuth angles of arrival (AoA),
$\bar{\theta}^{\rm tx}_\ell$, $\bar{\phi}^{\rm tx}_\ell$ are the average intra-cluster inclination and azimuth angles of departure (AoD).
These parameters are common across the clusters, reflecting that
the angles and delay are dependent on the large scale path routes
and do not depend on the frequency.
In contrast, the parameters 
 $F_{\ell m}$, $m=1,\ldots,M$, are the frequency-dependent components
given by
\begin{equation} \label{eq:Fell}
    F_{\ell m}
        = \left[\Delta \theta^{\rm rx}_{\ell m},
        \Delta \phi^{\rm rx}_{\ell m},
        \Delta \theta^{\rm tx}_{\ell m},
        \Delta \phi^{\rm tx}_{\ell m},
        \Delta \tau_{\ell m},
        P_{\ell m}
        \right],
\end{equation}
where $\Delta \theta^{\rm rx}_{\ell m}$, $\Delta \phi^{\rm rx}_{\ell m}$ are root mean square (RMS) inclination and azimuth angles of arrival spread, 
$\Delta \theta^{\rm tx}_{\ell m}$, $\Delta \phi^{\rm tx}_{\ell m}$ are the RMS inclination and azimuth angles of departure spread, 
$\Delta \tau_{\ell m}$ is the RMS delay spread,
$P_{\ell m}$ is the total received power.
The multi-frequency channel can then be described by the set of $L$ clusters
\begin{equation} \label{eq:theta}
    \nbx = \left[ \nbx^{(1)}, \ldots, \nbx^{(L)} \right].
\end{equation}
We will call $\nbx$ the \emph{cluster vector}.

To simplify the modeling, we will fix the number of clusters at $L=10$. When there
are less than $L$ clusters at a certain frequency, say $L_0 < L$, we simply set $P_{\ell m}=0$
(in linear scale) for all $\ell > L_0$ at frequency m.
Similarly, if a path cluster exist in one frequency $m$ but not 
at another frequency $m'$, we take $P_{\ell m} > 0$ and $P_{\ell m'} = 0$.
This scenario can occur, for example, where a path can penetrate materials
at a lower frequency, but be blocked at a millimeter wave frequency.

The above parameterization captures the fact that the geometric path route 
is independent of frequency.  However, the angular spread from scattering 
as well
as the transmission and reflection losses are well-known to be 
frequency dependent -- for example, see the scattering studies in THz
in \cite{degli2007measurement,schecklman2011terahertz,ju2019scattering}.
We also note that the 3GPP procedure for multi-frequency models
also generates paths with common cluster mean angles of arrival
and departure, but frequency dependent gains and 
angular spread \cite{3GPP38901}.  However, the frequency-dependent
components are assumed to be indepdent in the 3GPP model.  In contrast, 
our procedure is general and can, in principle, learn statistical relationships
with sufficient data.

Since the model \eqref{eq:thetaell} has $5+6M$
parameters per cluster, and $L$ clusters, there are a total of $(5+6M)L$ parameters in \eqref{eq:theta}. 
For example, in the example below, we will consider $M=2$ frequencies and $L=10$ clusters for a total of 
$p=(5+6M)L=170$ parameters.

Each link also has a  condition vector, $\nbu$.  
For our case, we will simply take 
\begin{equation} \label{eq:uvec}
    \nbu = (d_x,d_y,d_z),
\end{equation}
the 3D vector between the TX and RX.  Other parameters
could be added to $\nbu$ such as the antenna heights
or cell types.

The statistical channel modeling problem is to 
model the distribution of the cluster vector $\nbx$
as a function of the link conditions $\nbu$.  
That is, we want to model the conditional 
probability distribution $p(\nbx|\nbu)$.
In this work, we will be interested in so-called 
generative models where the conditional distribution
is represented as a mapping
\begin{equation} \label{eq:gen}
    \nbx = g(\nbu, \nbz),
\end{equation}
where $\nbz$ is some random vector with a known distribution.

Generative models \eqref{eq:gen} are the basis for
cellular evaluations such as \cite{3GPP38901}.
For example, the gNBs and UEs are first randomly 
placed in an environment following some deployment model
assumption (e.g.\ urban micro with some inter-site distance
and density).  The deployment determines the link conditions
$\nbu$ between each gNB and UE.  Using a generative model
\eqref{eq:gen}, we can then randomly generate path parameters
between all the gNBs and UEs.  Once we combine these
path parameters with the antenna array assumptions on both sides,
we can determine all the MIMO channels in the network.
One can then run any algorithm of interest to evaluate
relevant system performance statistics such as the SNR
or rate distribution.

\section{Urban Cellular Ray Tracing Data and Scattering Modeling}
\label{sec:raytracing}

\begin{figure}
  \centering
  \includegraphics[width=0.7 \linewidth]{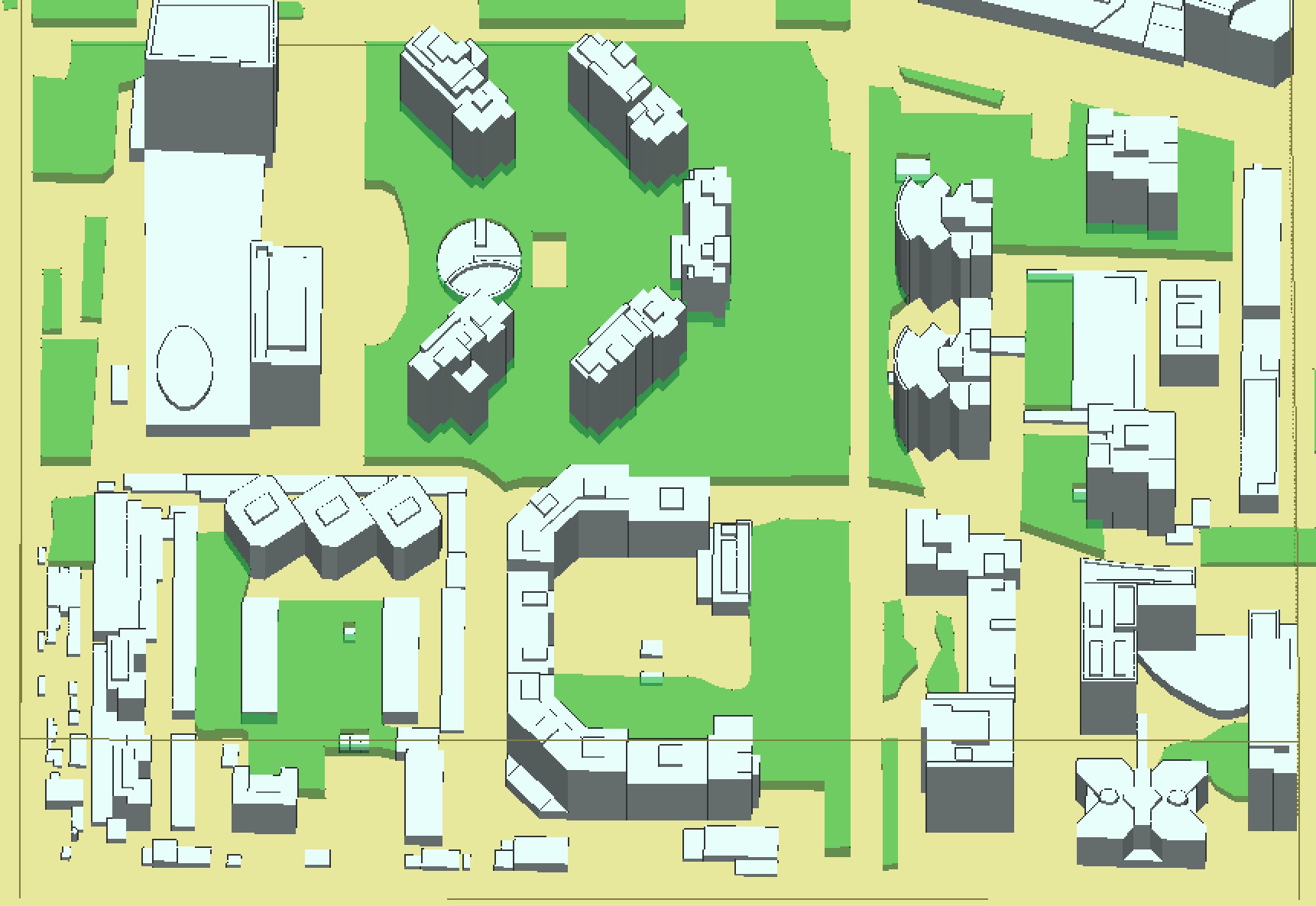}
  \caption{Ray tracing simulation area representing a 550 $\times$ 380 $\text{m}^2$ region of Beijing, China.}
  \label{fig:beijing_map}
\end{figure}

Although our methodology is general,
we will demonstrate the methods on two frequencies:  $f_1=$\, \SI{28}{GHz}, 
as used in 5G systems today \cite{shafi20175g}, 
and $f_2=$\, \SI{140}{GHz}, a key sub-THz frequency targeted for 
potential 6G use cases \cite{giordani2020toward,rappaport2019wireless}.
While there have
been recent multi-frequency measurement
campaigns
\cite{huang2020multi,cui2020multi},
these measurements are 
enormously time-consuming and there are no current
datasets with sufficient data points for
training complex neural networks.
Thus, similar to several recent works \cite{XiaRanMez2020,Alkhateeb2019,khawaja2017uav}
we use a powerful ray tracing tool, Wireless InSite by Remcom \cite{Remcom} to generate data for training. 
\iftoggle{conference}{
Specifically, an area of Beijing, China, consisting of 550 $\times$ 380 square meters as shown in Fig.~\ref{fig:beijing_map}, is imported. Using similar parameters to 3GPP report \cite{3GPP38901}, we manually place 100 terrestrial receiving gNBs on the outside of buildings with \SI{10}{m} height, 
and transmitting UEs are placed at 320 locations on streets \SI{1.5}{m}. 
The deployment results in 100 $\times$ 320 $=$ 32000 links (i.e. gNB-UE pairs).
We run independent ray tracing simulations 
on these the two frequencies.
We set the maximum number of reflections as six and the maximum number of diffraction as one, and we limited the maximum number of paths of a link to 20.
In addition, the ray tracing tool only collects paths with a power of at least -250 dBm.
}
{
An area of Beijing, China, consisting of 1650 $\times$ 1440 square meters, has been imported. This representation of Beijing includes building and terrain data, as shown in Fig.~\ref{fig:beijing_map}.
To generate the data set, we manually placed 180 terrestrial receiving gNBs on the outside of the building with \SI{2}{m} height, imitating the typical locations of 5G PicoCell stations which are designed to serve the ground users. 
The transmitting UEs were placed at 120 locations on streets \SI{1.5}{m} high resembling street-level users.  These numbers are also consistent
with micro-cellular models in \cite{3GPP38901}.  

With the above gNBs and UEs configuration, there are total of 180 $\times$ 120 $=$ 21600 links (i.e. gNB-UE pairs). 
We ran independent ray tracing simulations 
on these links at
two frequencies:  $f_1=$\, \SI{2.3}{GHz}
and $f_2=$\, \SI{28}{GHz} -- typical values
for a sub-6 GHz and mmWave carrier
in commercial deployments today.

The materials will have different 
permittivity and conductivity values at different
frequency.  In this simulation, we use the values Table~\ref{table:material_para}. 
Furthermore, we set the maximum number of reflections as six and the maximum number of diffraction as one, and we limited the maximum number of paths of a link to 20. 

\begin{table}
\caption{Material Parameters$^{*}$}
\begin{center}
\begin{tabular}{c|c|c|c|c|}
\cline{2-5}
                                         & \multicolumn{2}{c|}{Earth Ground} & \multicolumn{2}{c|}{Buliding Wall} \\ \hline
\multicolumn{1}{|c|}{Frequency (GHz)}     & 2.3              & 28             & 2.3              & 28              \\ \hline
\multicolumn{1}{|c|}{Permittivity (F/m)} & 21.13            & 5.70           & 5.31             & 5.70            \\ \hline
\multicolumn{1}{|c|}{Conductivity (S/m)} & 0.47             & 9.50           & 0.07             & 9.50            \\ \hline
\multicolumn{4}{l}{$^{*}$Default value in Remcom database \cite{Remcom}}
\end{tabular}
\end{center} 
\label{table:material_para}
\end{table}
}

Critical to modeling the frequency dependence is the effect of diffuse
scattering, which is particularly prominent in the mmWave and THz
frequencies \cite{schecklman2011terahertz,ju2019scattering,xing2021millimeter}.
As part of our simulation, we enable the diffusing scattering model in the ray tracing tool which is based on the directive model developed by \cite{degli2007measurement}. A key parameter is the scattering
coefficient, $S$, which is the ratio of the scattered to incident
electric field.
Following measurements in  \cite{degli2007measurement}, 
we set the scattering coefficient 
 to $S=0.273$ for \SI{28}{GHz}, and $S=0.4$ for \SI{140}{GHz},
in which case, rays in \SI{140}{GHz} disperse more power on the forward lobe of diffuse scattered.
A diffuse scattering interaction is limited to the last interaction along a path before it reaches the receiver. This helps reduce the run time for diffuse scattering calculations. 

\section{Paths Cluster and Generative Neural Network Modeling}
\label{sec:gan}


\begin{figure*}[t]
\footnotesize
\centering
\includegraphics[width=0.95 \linewidth]{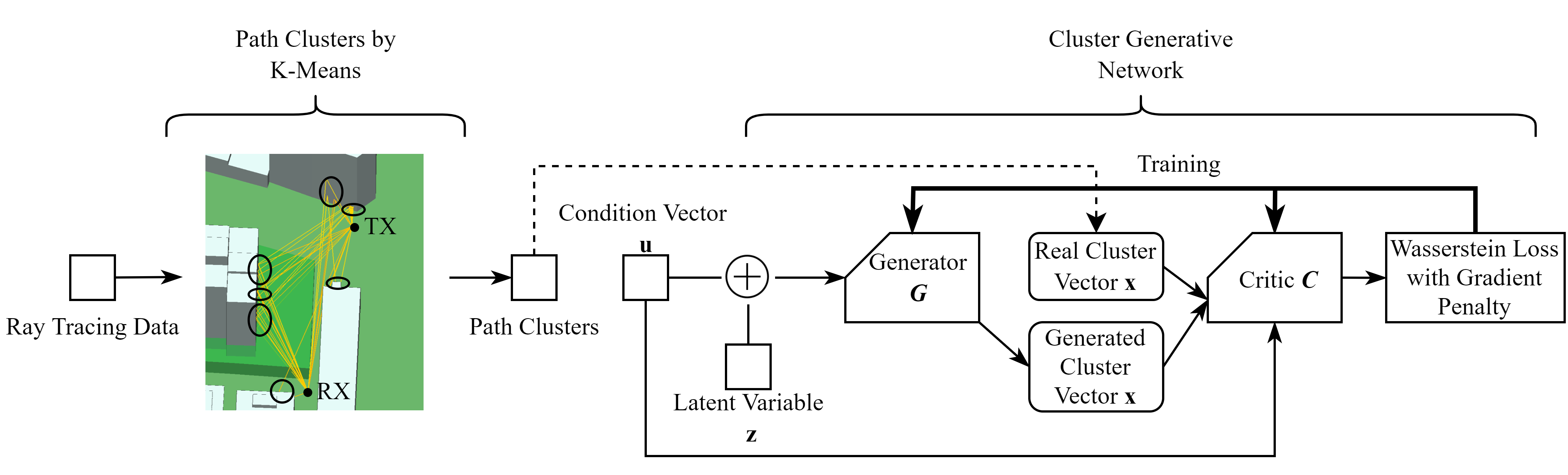}
\caption{The architecture for the K-Means path clusters generator and the generative adversarial neural network model}
\label{fig:gen_model}
\vspace{-2mm}
\end{figure*}

\subsection{Path Clusters Identification by $K$-means}
The ray tracing data produces rays at each frequency.
Our first step is to pre-preocess the rays to identify the 
clusters and determine the cluster parameters.
Similar to \cite{he2018clustering}, we use a 
$k$-means clustering procedure 
on the inclination and azimuth angles of arrival, inclination and azimuth angles of departure, and delay for each link of the set. Silhouette scores \cite{rousseeuw1987silhouettes} are used 
to determine the best number of clusters for each link.
A typical identification of the clusters is shown in the map in 
\eqref{fig:gen_model} where the circles indicate the identified clusters.

After identifying the clusters, we can determine the angular and delay spread
and power in each cluster to to generate the cluster vector $\nbx$ as indicated in \eqref{eq:theta} for each link.
To limit the size of the cluster vector, we use $L=10$ as the maximum number of clusters.

We obtain one data record $\nbx$ for each link.  Each link also has 
condition vector $\nbu$ as in \eqref{eq:uvec}.
The condition vector $\nbu$ is first converted to a 3D distance and the logarithm of the 3D distance, denoted $\nbu=(d_{3D}, \log_{10}d_{3D})$.

\iftoggle{conference}{
}
{ 
In which case, the output of link state network $s$, as shown in Fig. \ref{fig:gen_model}, 
is denoted as $s \in \{\verb|LOS|,\verb|NLOS|,\verb|Outage|\}$. 
We design a fully connected neural network whose input is the condition vector $\nbu$, and it generates the probabilities of the three link states.
The parameters of this neural network is shown in Table \ref{table:model_config}. 
Specifically, the condition vector $\nbu$ is transformed to the new vector,
\begin{equation} \label{eq:transformed_cond_vector}
    \nbu = (d_{3D},\log_{10}d_{3D}),
\end{equation}
where $d_z$ is the z-axis (vertical) distance between the TX and RX, and the 3D distance value $d_{3D}$ is denoted as:
\begin{equation} 
    d_{3D} = \sqrt{d_x^2 + d_y^2 + d_z^2}.
\end{equation}
We use a standard scaler to process the transformed condition vector in \eqref{eq:transformed_cond_vector}, 
then through two hidden layers, a three ways soft-max activation function is used to obtain the probabilities of the three link states. 
In the end, we use a uniform random variable $r.v. \in [0,1]$ to sample the soft-max result and get the output of link state network $s$.
}

\iftoggle{conference}{
\subsection{Cluster Generative Network} 
The goal of the network is to generate random cluster vectors $\nbx$ in \eqref{eq:theta} following the conditional distribution $p(\nbx|\nbu)$ as observed in the data.  
For this purpose, we use a conditional Wasserstein generative adversarial network with gradient penalty (CWGAN-GP) \cite{gulrajani2017improved,arjovsky2017wasserstein} that is relatively simple to implement while being effective in complex datasets.

The CWGAN-GP can be summarized as follows:  There are two components:  
a \emph{generator} $\nbG$
and \emph{critic} $\nbC$.  The generator takes the condition vector $\nbu$ and some
random input $\nbz \sim p(\nbz)$ and generates a random path data vector $\nbx = \nbG(\nbu,\nbz)$.
The vector $\nbz$ is called the \emph{latent vector}.
As is commonly used, we take its distribution $p(\nbz)$ to be a unit variance Gaussian vector.  We set the latent dimension to 20.
To ensure the generator matches the data distribution, one also trains a critic
function $\nbC$ that attempt to discriminate between the generated and true samples.
The critic $\nbC$ and generator $\nbG$ networks are trained via a loss:
\begin{align}
    \label{eq:wasser_loss}
    L= & \underbrace{\underset{\tilde{\boldsymbol{x}} \sim \mathbb{P}_{g}}{\mathbb{E}}[C(\tilde{\boldsymbol{x}})]-\underset{\boldsymbol{x} \sim \mathbb{P}_{r}}{\mathbb{E}}[C(\boldsymbol{x})]}_{\text {Original critic loss }}+ \nonumber \\
    & \underbrace{\lambda \underset{\hat{\boldsymbol{x}} \sim \mathbb{P}_{\hat{\boldsymbol{x}}}}{\mathbb{E}}\left[\left(\left\|\nabla_{\hat{\boldsymbol{x}}} D(\hat{\boldsymbol{x}})\right\|_{2}-1\right)^{2}\right] .}_{\text {Gradient penalty }}
\end{align} 
where  $\mathbb{P}_g$ is the generator distribution implicitly defined by $\tilde{\boldsymbol{x}}=\nbG(\boldsymbol{u},\boldsymbol{z}), \quad \boldsymbol{z} \sim p(\boldsymbol{z})$,
and $\mathbb{P}_r$ is the distribution of the data points.  The loss \eqref{eq:wasser_loss}
is optimized via a minimax:  The critic $\nbC$ attempts to minimize the loss
to discriminate between the generated and true samples, while the 
generators attempts to maximize the loss to fake the critic.  The minimax is similar to other GANs,
but the key concept in the WGAN-GP is that the gradient penalty term avoids mode collapse
-- see \cite{gulrajani2017improved,arjovsky2017wasserstein} for details.  
For our application, both the generator and critic are realized with as fully connected
neural networks with parameters shown in Table \ref{table:model_config}.

One important detail for the wireless channel modeling is that the
condition vector $\nbu$ and cluster data $\nbx$ have heterogeneous data types and need
to be brought to a uniform format.  
To overcome this challenge, we align the angles of each path to the LOS direction and measure excess signal delay relative to the LOS distance delay. 
We then scale all the values using either min-max or standard scalers.  
Details are in the code~\cite{mmwchanmod-GAN}.
}
{
\subsection{Cluster Generative Network}
\textcolor{red}{Change path to cluster}
The path generative network is implemented by a conditional Wasserstein generative adversarial network with gradient penalty (CWGAN-GP) \cite{gulrajani2017improved} \cite{arjovsky2017wasserstein}. 
Its goal is to create imitative parameters of cluster vector in \eqref{eq:theta} based on the training data.
The CWGAN-GP is constructed by a generator $\nbG$ and a critic $\nbC$ as shown in Fig. \ref{fig:gen_model}. 
The generator $\nbG$ uses the condition $\nbc$ and the latent variable $\nbz$ to generate the imitative cluster vector $\nbx$.
And the critic $\nbC$ generates the Wasserstein loss based on criticizing the real cluster vector, the generated cluster vector, and the generative adversarial network condition $\nbc$.
In the end, we train the whole path generative network by the Wasserstein loss with gradient penalty, which is going to be explained more in below.

\subsubsection{Generator inputs transformation}
There are two parts of the input of the generator $\nbG$, 
the latent variable $\nbz$ and the GAN condition $\nbc$, as shown in Fig. \ref{fig:gen_model}. 
The GAN condition $\nbc$ is transformed by the condition vector $\nbu$ and the link state $s$, 
\begin{equation} \label{eq:GAN_cond_vector}
    \nbc = (\nbu, s),
\end{equation}
and we use a standard scaler to process it. 
And in our \SI{2.3}{GHz} and \SI{28}{GHz} experiment, a 20-dimensional Gaussian vector $\nbz$, which $z \sim \mathcal{N}(0,\,1)$, is used. 
Thus, the input of the generator $\nbG$ is denoted by $((\nbu, s)_{\text{scaled}}, \nbz)$

\subsubsection{Critic inputs transformation}
The inputs of critic $\nbC$ are the GAN condition $\nbc$ and the cluster vector $\nbx$. 
As same as transforming in the generator, a standard scaler is used to process $\nbc$. 
As we mentioned in Table \ref{table:model_config}, in our example, a UE-gNB link cluster vector $\nbx$ includes 140 parameters.
Specifically, these parameters are heterogeneous and we need to scale them in different ways to put them on an equal footing.

The transformation detail shows in the following:
\begin{enumerate}
    \item Delay: 
    The LOS delay represents the distance between the TX and RX divided by speed of light. 
    Then the delay of paths subtract by the LOS delay. 
    And a Min-Max scaler is used to scale the resulting excess path delays;
    \item Azimuth angles: 
    The azimuth angle of arrival and departure are rotated relative to the LOS angle which is the relative angel between TX and RX. 
    Then these two angle are scaled into $[-1,1]$ by dividing 180\degree;
    \item Elevation angles: 
    The elevation angle of arrival and departure are first rotated relative to LOS angle. 
    Then we use a standard scalar which normalizes by the standard deviation;
    \item Path gains:
    From equation \eqref{eq:thetaell}, a path $\theta$ has path gains $G_{1},\ldots,G_{M}$ across the $M$ frequencies. 
    To scale the path gains properly, the free space path losses $FSPL_{1},\ldots,FSPL_{M}$ across multiple frequencies are calculated by the Friis' Law, 
    $FSPL \text{(dB)} = 20\log_{10}(4\pi d f /c)$, where $c$ is the light speed, d is the distance between the TX and RX, and f is the frequency.
    We scale the path gain $G_{k}$ to 0 if the negative value of $G_{k} \text{(dB)}$ is $\Delta$ dB larger than the free space path loss $FSPL_{k} \text{(dB)}$ in corresponding frequency $k$.
    For other path gains, we use a excess value scaling method which is
    calculating the excess value $e_k$ by subtracting the free space path loss $FSPL_{k} \text{(dB)}$ from the negative value of $G_{k} \text{(dB)}$ in each frequency $k$, 
    and then the results transform to $[0,1]$ by $(\Delta - e_k)/\Delta$.
    After this transformation, the large gains are scaled closed to 1 and small gains are closed to 0,
    and we use $\Delta = 80 \text{dB}$ in our \SI{2.3}{GHz} and \SI{28}{GHz} example.
\end{enumerate}
These transformations make sure that different classes of values scale into a similar range and reference to the LOS path.
The Min-Max scaler of delay is fit to the training data.
We note that the rotation of angles helps generative network generates the angles of NLOS paths.
And it is important to use a excess value of the free space path loss scaling method on the path gains across multiple frequencies, which has a positive effect on training the critic $\nbC$.

\subsubsection{Conditional WGAN-GP model}
A conditional WGAN-GP model, as shown in Table \ref{table:model_config}, is used to generate the cluster vector parameters $\nbx$ in \eqref{eq:theta}. According to a specific GAN condition $\nbc$, the generator $\nbG$ produces the counterfeit cluster vector $\nbx$ and the critic $\nbC$ discriminates the real cluster vector and the generated fake cluster vector. Based on the appraisal results of the critic, we implement a wasserstein loss with gradient penalty as
\begin{align}
    \label{eq:wasser_loss}
    L= & \underbrace{\underset{\tilde{\boldsymbol{x}} \sim \mathbb{P}_{g}}{\mathbb{E}}[D(\tilde{\boldsymbol{x}})]-\underset{\boldsymbol{x} \sim \mathbb{P}_{r}}{\mathbb{E}}[D(\boldsymbol{x})]}_{\text {Original critic loss }}+ \nonumber \\
    & \underbrace{\lambda \underset{\hat{\boldsymbol{x}} \sim \mathbb{P}_{\hat{\boldsymbol{x}}}}{\mathbb{E}}\left[\left(\left\|\nabla_{\hat{\boldsymbol{x}}} D(\hat{\boldsymbol{x}})\right\|_{2}-1\right)^{2}\right] .}_{\text {Gradient penalty }}
\end{align} 
where $D$ is the discriminator and $\mathbb{P}_g$ is the generator distribution implicitly defined by $\tilde{\boldsymbol{x}}=\nbG(\boldsymbol{c},\boldsymbol{z}), \quad \boldsymbol{z} \sim p(\boldsymbol{z})$.
In our case, since the discriminator $D$ in WGAN is not trained to classify, the $D(\boldsymbol{x})$ here actually is implemented by the critic $\nbC(\boldsymbol{x})$.
We implicitly define $\mathbb{P}_{\hat{\boldsymbol{x}}}$ sampling uniformly along straight lines between
pairs of points sampled from the data distribution $\mathbb{P}_{r}$ \cite{gulrajani2017improved} \cite{arjovsky2017wasserstein}. Also we use the recommend value of $\lambda = 10$ in paper \cite{gulrajani2017improved}.

The number of parameters of the path generative network, as shown in Table \ref{table:model_config}, are different from the traditional dense layer parameter calculation method.
For the path GAN generator, before merging two inputs (20-dimensional latent variable $\nbz$ and 3-dimensional GAN condition $\nbc$), 
two dense layers with 280 neurons are respectively used for two inputs.
After merging, a Batch-Normalization layer is used, whose parameters are not trainable.
For the path GAN critic, the 3-dimensional GAN condition $\nbc$ expands to 140-dimensional by using a 140 units dense layer, and then it is merged with another input that is the 140-dimensional cluster vector $\nbx$.
}

\begin{table}[t]
\caption{Generative model configuration}
\begin{center}
\begin{tabular}{l|c|c|}
\cline{2-3}
                                         & \begin{tabular}[c]{@{}c@{}}Cluster GAN \\ Critic\end{tabular} & \begin{tabular}[c]{@{}c@{}}Cluster GAN \\ Generator\end{tabular} \\ \hline
\multicolumn{1}{|l|}{Num. of inputs}     & 170 $+$ 2                                                & 20 $+$ 2                                                      \\ \hline
\multicolumn{1}{|l|}{Hidden units}       & {[}1120,560,280{]}                                         & {[}280,560,1120{]}                                            \\ \hline
\multicolumn{1}{|l|}{Num. of outputs}    & 1                                                          & 140                                                           \\ \hline
\multicolumn{1}{|l|}{Optimizer}          & \multicolumn{2}{c|}{Adam}                                                                                                  \\ \hline
\multicolumn{1}{|l|}{Learning rate}      & \multicolumn{2}{c|}{0.0001}                                                                                                \\ \hline
\multicolumn{1}{|l|}{Epochs}             & \multicolumn{2}{c|}{10000}                                                                                                  \\ \hline
\multicolumn{1}{|l|}{Batch size}         & \multicolumn{2}{c|}{1024}                                                                                                  \\ \hline
\multicolumn{1}{|l|}{Num. of parameters} & 1167551                                                    & 1142290*                                                       \\ \hline
\multicolumn{3}{l}{$^{*}$Different from the traditional parameters calculation method}
\end{tabular} 

\end{center}
\label{table:model_config}
\end{table}
\section{Modeling Results}
\label{sec:modeling_results}

As described in Section~\ref{sec:raytracing}, the data consists of 32000 links,
with each link having the paths at both \SI{28}{GHz} and \SI{140}{GHz}.  
We show the path clusters following on a GAN framework is able to capture interesting joint statistics between these two frequencies.  
In all the experiments below, we use 80\% of the links for training and 20\% for test.  
The generative network is implemented in Tensorflow 2, and the data set, code, and pre-train models can be accessed through \cite{mmwchanmod-GAN}.

\subsection{Received Power with Omnidirectional Antennas}
\begin{figure}
  \centering
  \includegraphics[width=1.0 \linewidth]{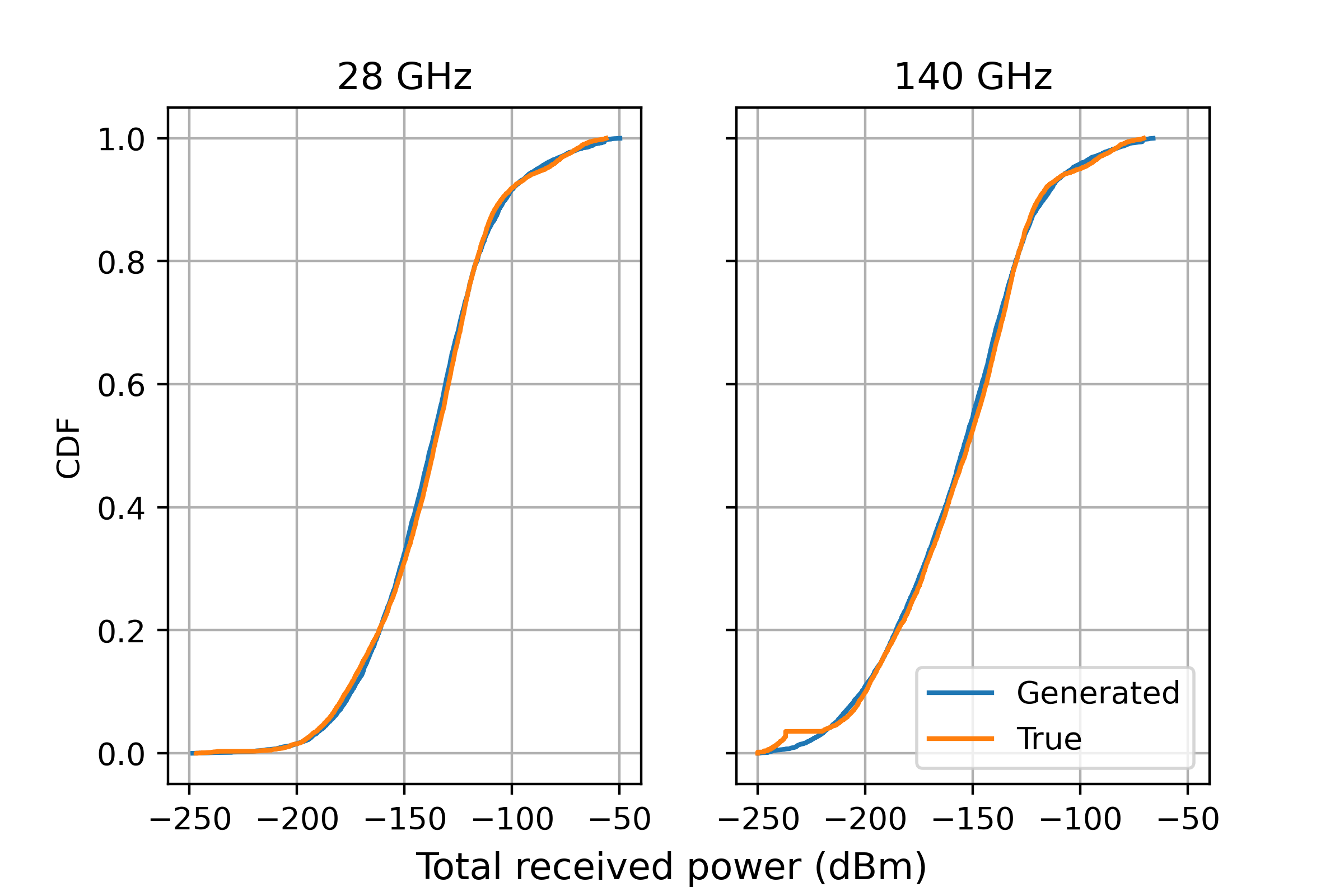}
  \caption{The CDF of the total received power of links on the test data versus the CDF of randomly generated links' total received power from the trained model.}
  \label{fig:received_power_cdf}
\end{figure}

\begin{figure}
  \centering
  \includegraphics[width=1.0 \linewidth]{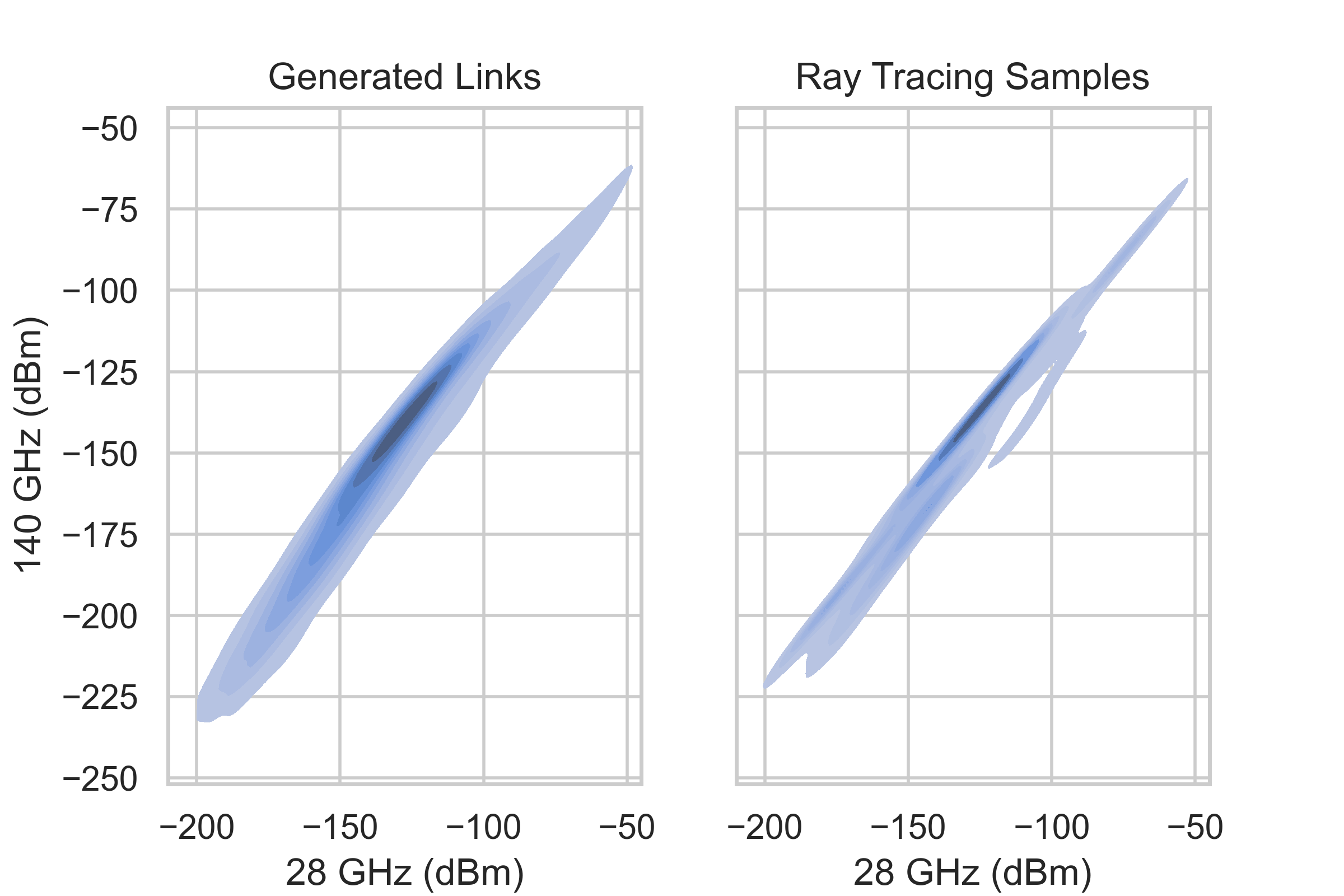}
  \caption{The KDE plot of bivariate distributions of the total received power in \SI{28}{GHz} and \SI{140}{GHz}.}
  \label{fig:received_power_comparing}
\end{figure}

We first evaluate the ability of the model to capture the joint distribution of the total received power at the two frequencies.  
The comparison can be performed as follows:
Let $\tilde{\boldsymbol{x}}=\nbG(\boldsymbol{u},\boldsymbol{z})$ denote the learned generative model
and let $(\nbu_i,\nbx_i)$, $i=1,\ldots,N_{\rm ts}$ denote the test data,
where each sample contains the link condition $\nbu_i$ 
and the corresponding cluster data vector $\nbx_i$.  For each test sample, we can compute the vector 
\begin{equation} \label{eq:vdata}
    \nbv_i = (v_{i1},v_{i2}) = (\phi_1(\nbx_1), \phi_2(\nbx_2)),
\end{equation}
where $v_{ij}$ is the omni-directional received power on sample $i$ at frequency $j$ where $j=1,2$.
Here, $\phi_j(\nbx)$ is the function that computes the omni-directional received power at frequency $j$ from the cluster data $\nbx$.
The omni-directional received power is simply the received power that would be experienced if the gNB and UE had omni-directional antennas.  

To compare these values with the generated distribution, for each test sample $i$,
we also generate a random sample $\nbx_i^{\text{rnd}} = \nbG(\nbu_i,\nbz_i)$ using the trained 
generator $\nbG$ and a random $\nbz_i$ with the same conditions $\nbu_i$ as the test data. 
We can then compute a set of generated received powers 
\begin{equation} \label{eq:vgen}
    \nbv_i^{\rm rnd} = (v_{i1}^{\rm rnd},v_{i2}^{\rm rnd}) = (\phi_1(\nbx_1^{\rm rnd}), \phi_2(\nbx_2^{\rm rnd})),
\end{equation}
Ideally, the samples $\nbv_i$ and $\nbv_i^{\rm rnd}$ should have a similar distribution.

Fig.~\ref{fig:received_power_cdf} plots the empirical cumulative distribution functions (CDFs) for the \emph{marginal} distributions for the data and model
at the two frequencies.  Specifically, the left plot shows the CDF of the data $v_{ij}$ and generated 
samples $v_{ij}^{\rm rnd}$ at frequency $j=1$.  The right plot shows the data and generated CDF for frequency $j=2$.
We see in both cases, these two marginal CDFs match almost perfectly.

But, what is most interesting is that the generator can also capture the joint statistics.  The left panel of Fig.~\ref{fig:received_power_comparing} shows a kernel density estimation (KDE) plot of the test data points $(v_{i1},v_{i2})$
and the right panel shows a KDE  plot of the trained model $(v_{i1}^{\rm rnd},v_{i2}^{\rm rnd})$.
We see that the two \emph{joint} distributions match well.

\subsection{Delay Distribution and RMS Delay Spread}
\begin{figure}
  \centering
  \includegraphics[width=0.9 \linewidth]{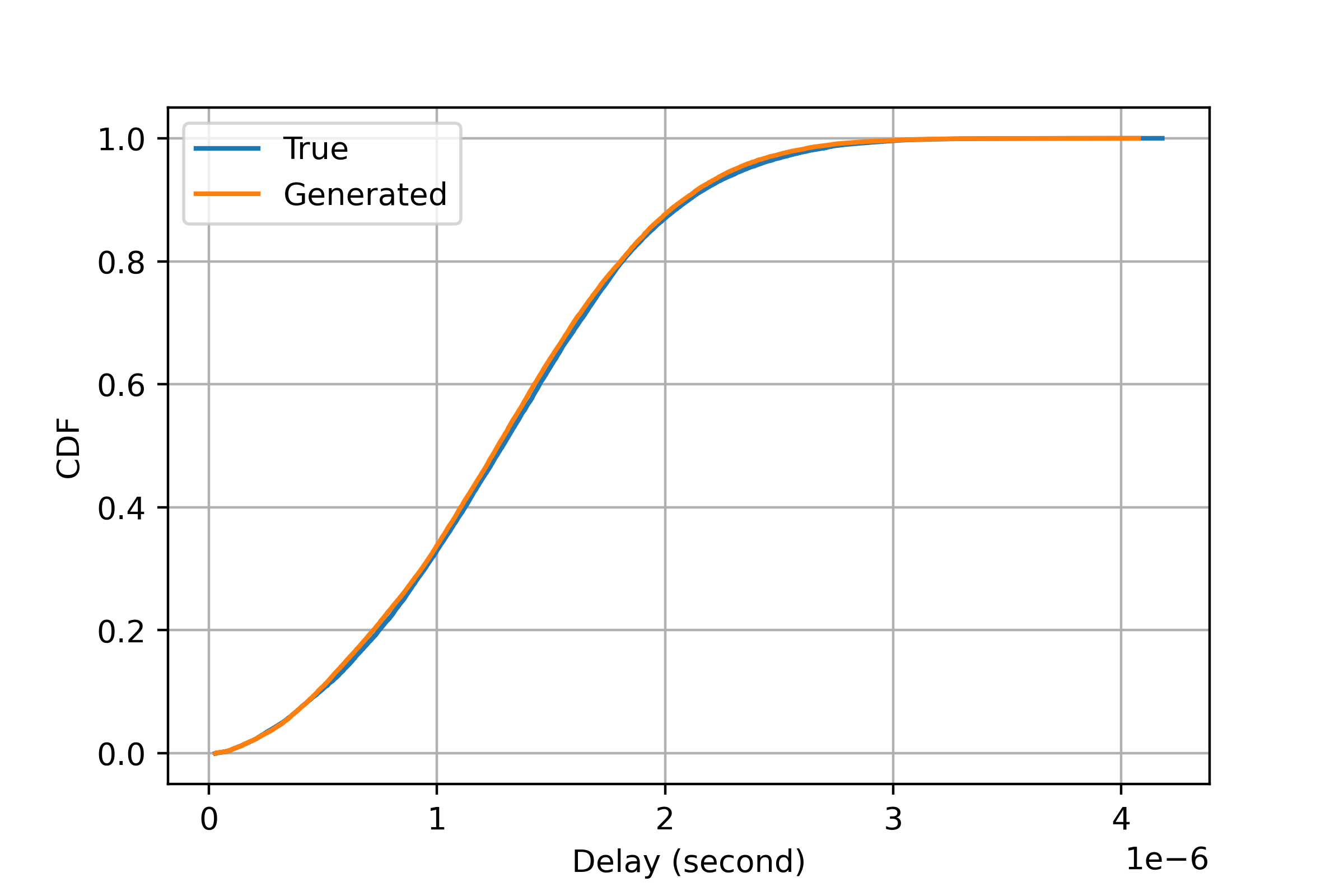}
  \caption{The CDF of the paths minimum delay in clusters.}
  \label{fig:delay_cdf}
\end{figure}
\begin{figure}
  \centering
  \includegraphics[width=0.9 \linewidth]{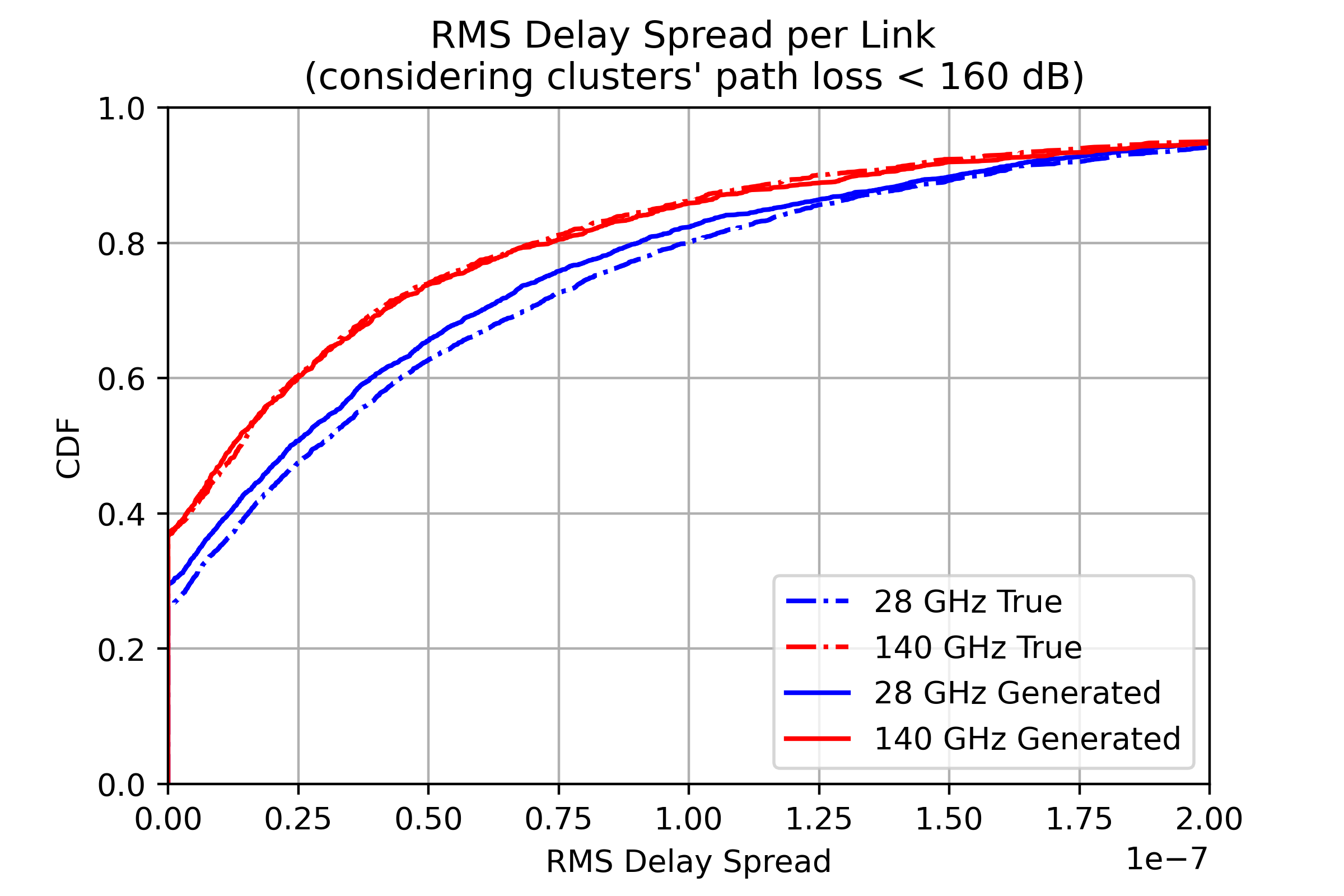}
  \caption{The CDF of the RMS cluster delay spread per link, where we only considering the path loss of the clusters are smaller than 160 dB.}
  \label{fig:delay_rms_cdf}
\end{figure}

To evaluate the model's ability to generate the delay, we first plot the CDF of the minimum path delay over the clusters, as shown in Fig.~\ref{fig:delay_cdf}.
There is an almost identical delay between clusters generated by the network and the real ray tracing value, and the maximum delay is also around 4000 nanoseconds.
For each link, we also compute the RMS delay spread in order to gain an intuitive understanding of how the model fits the delays across the different clusters within a link.
Fig.~\ref{fig:delay_rms_cdf} shows that the RMS delay spread at \SI{140}{GHz} is smaller than at \SI{28}{GHz} overall.
As the frequency increases, the RMS delay spread becomes smaller due to the increase in power loss by diffuse scattering. This conclusion is consistent with the mmWave and sub-Terahertz measurements in \cite{xing2021millimeter}.

\subsection{Angles Distribution and RMS Angles Spread}
\begin{figure}
  \centering
  \includegraphics[width=0.9 \linewidth]{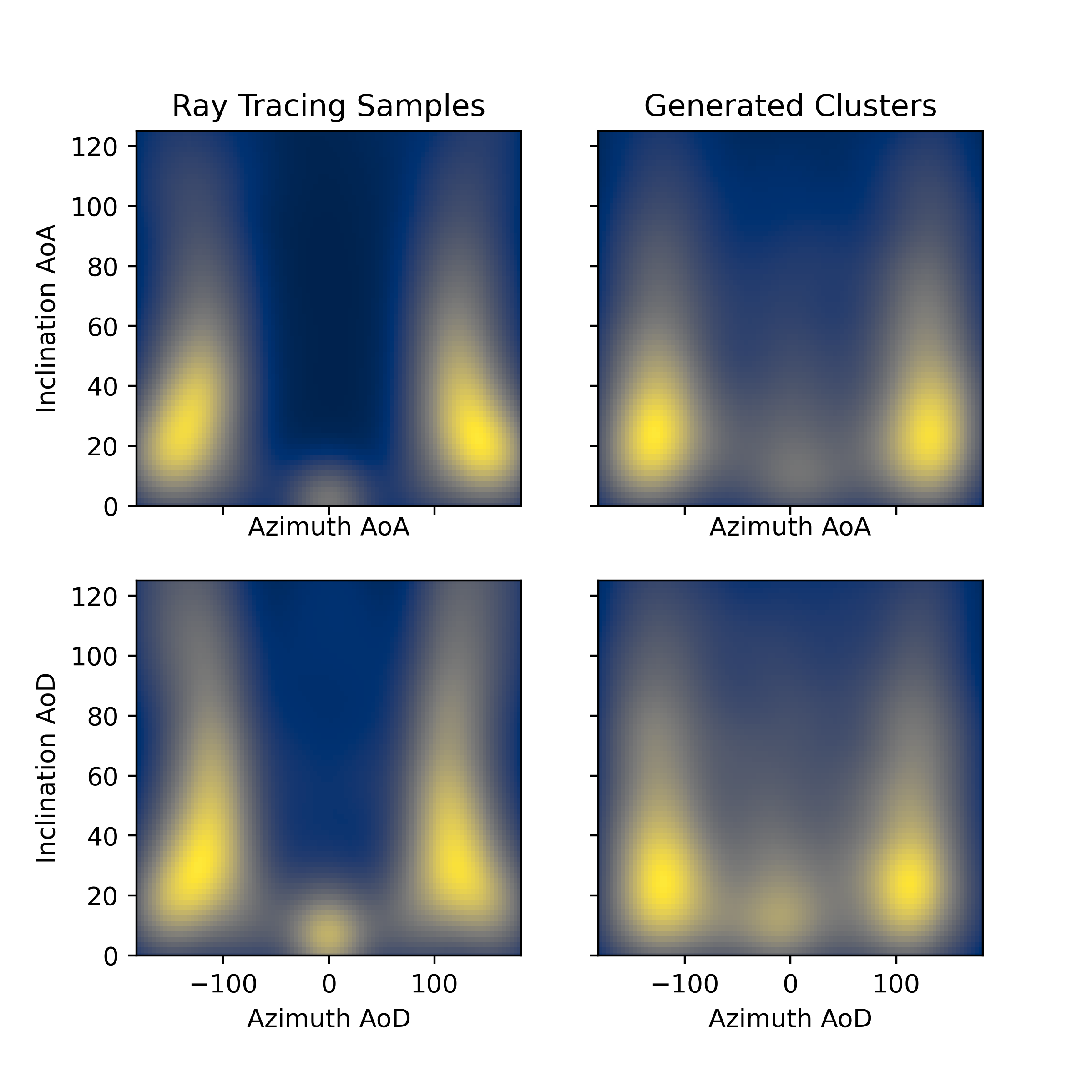}
  \caption{Azimuth angles vs. inclination angles, where we only consider the path loss of the clusters that are smaller than 160 dB.}
  \label{fig:az_vs_inc}
\end{figure}

\begin{figure}
  \centering
  \includegraphics[width=0.9 \linewidth]{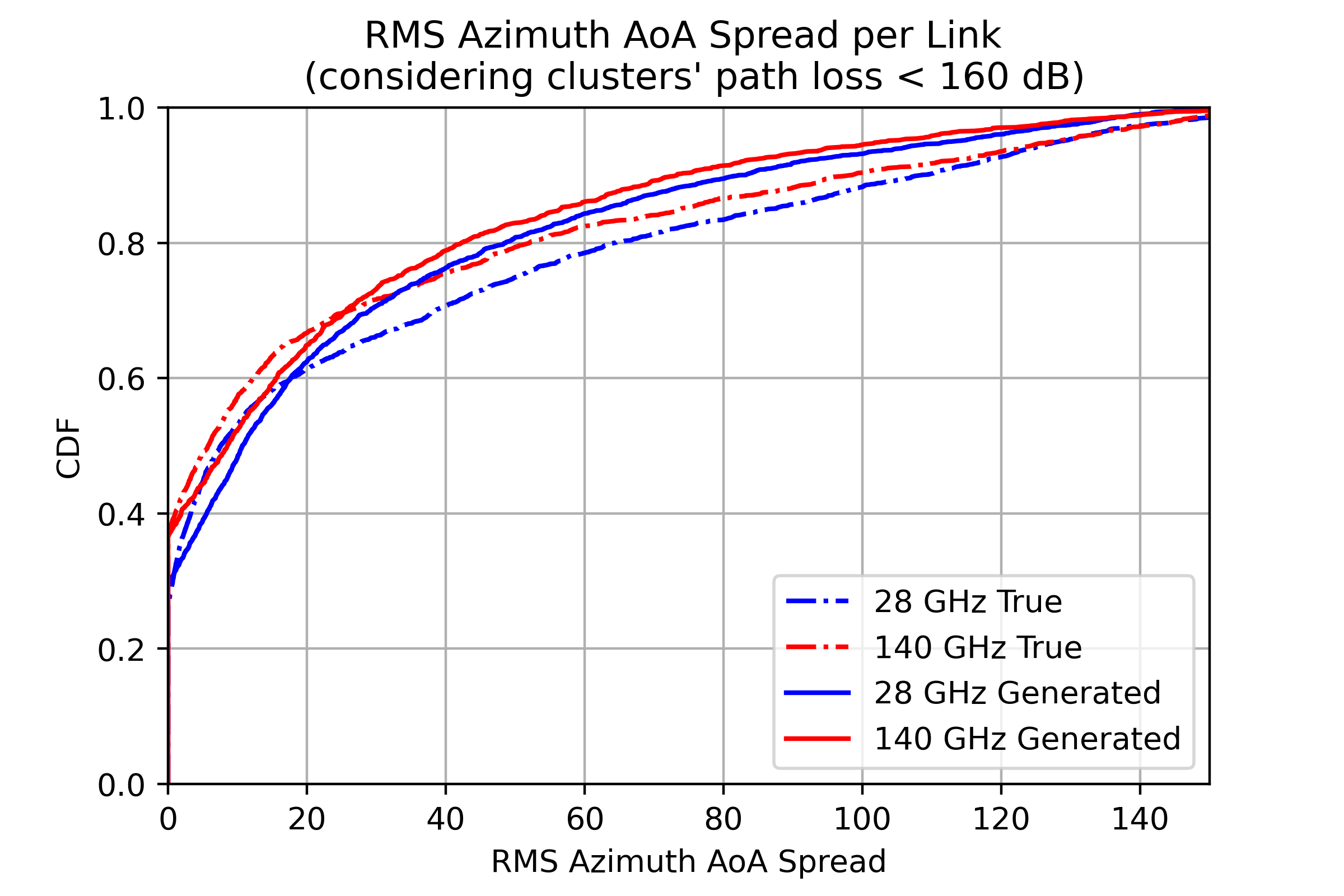}
  \caption{RMS azimuth AoA spread, including the zero values.}
  \label{fig:rms_aoa_az}
\end{figure}
\begin{figure}
  \centering
  \includegraphics[width=0.9 \linewidth]{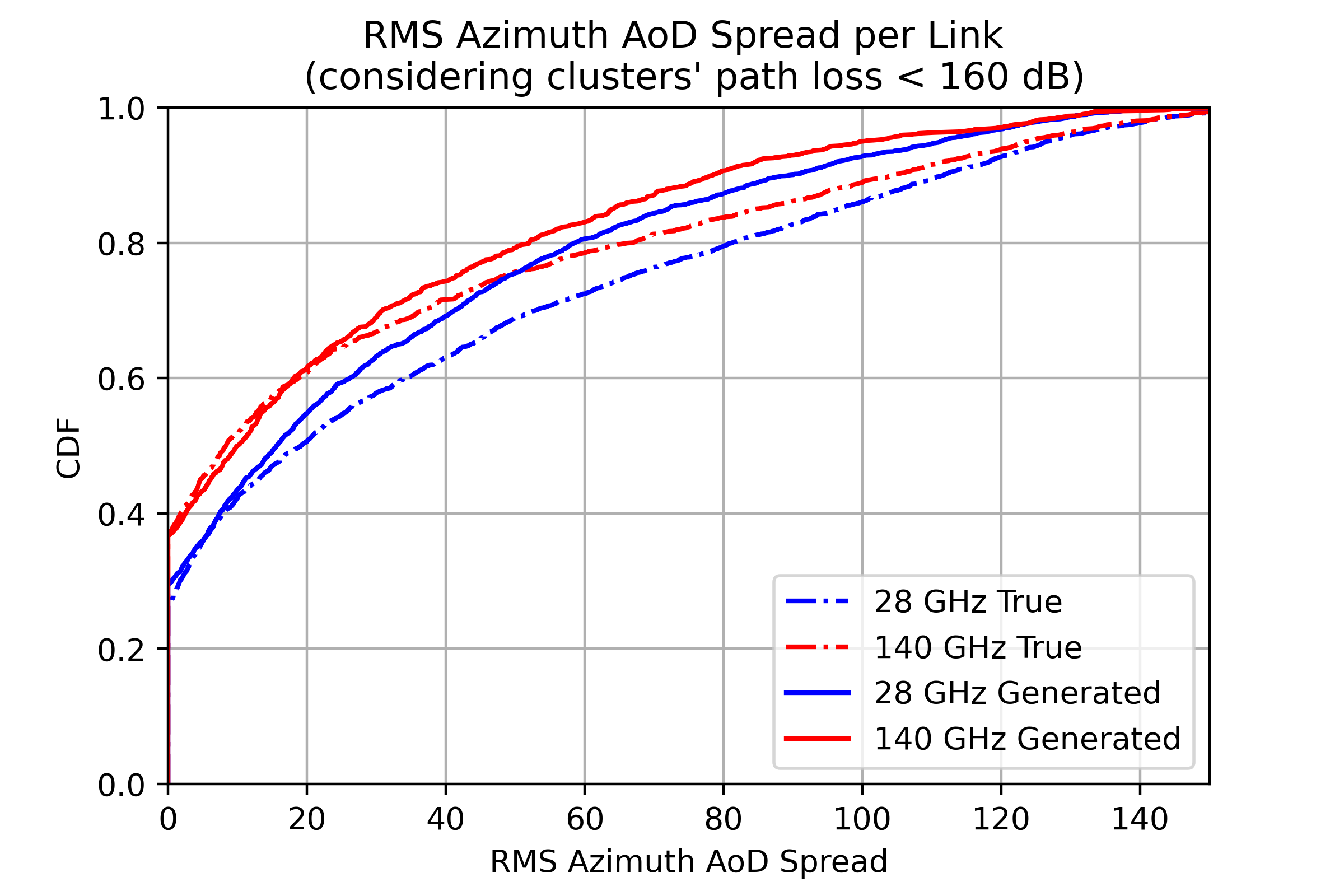}
  \caption{RMS azimuth AoD spread, including the zero values.}
  \label{fig:rms_aod_az}
\end{figure}

As we mentioned above, to overcome the challenge of heterogeneous data types, we align the angles of each path to the direction of LOS.
More precisely, when preprocessing the angles of arrival, we choose the LOS AoA direction of a link as the z-axis of the new spherical coordinate system and calculate the associated azimuth and inclination transformation angles. 
Similarly, in the case of angles of departure, the AoD direction of LOS is used.
In this way, the network can more effectively learn a statistical model of the angle relative to the condition vector.
As illustrated in Fig.\ref{fig:az_vs_inc}, both AoA and AoD distributions are similar to those observed in real ray tracing samples.
On the other hand, the RMS angle spread has important implications for evaluating channel models, especially when considering mmWave and sub-Terahertz bands.
Fig.\ref{fig:rms_aoa_az} and Fig.\ref{fig:rms_aod_az} show CDF images of AoA and AoD RMS spread, respectively.
Overall, the model is able to capture the properties of the RMS angle spread at both frequencies.
As a result of our testing, the RMS angle spread at \SI{28}{GHz} is slightly greater than at \SI{140}{GHz}. 
When there is only one cluster on a link at a given frequency, the RMS angle spread is zero. 
Based on Figs.~\ref{fig:rms_aoa_az} and \ref{fig:rms_aod_az}, we can see that at \SI{140}{GHz}, the proportion of RMS spread value of zero is higher than that at \SI{28}{GHz}, owing to higher diffuse scattered power and increased path loss, which reduces the number of paths that can be detected by the receiver.
The above observations are consistent with the conclusions obtained from actual measurements at \SI{28}{GHz} and \SI{140}{GHz} in \cite{xing2021millimeter}.

\section{Conclusion}
\label{sec:conclusion}
Wireless multi-path channels can exhibit complex statistical relationships 
across different frequencies which are difficult to capture via standard models such as \cite{3GPP38901}, particularly for mmWave and THz channels.
In this work, we have presented a general modeling methodology for deriving full double
directional models at multiple frequencies.  The methods use state-of-the-art neural networks
and make minimal a priori assumptions in the data.  As such, we show they are able to capture
interesting cross-frequency relations.  The current method is based on extensive ray tracing data. 
A natural extension, left as future work, will be to augment the data with real data as these become available.

\bibliographystyle{IEEEtran}
\bibliography{bibl}

\begin{thebibliography}{10}
\providecommand{\url}[1]{#1}
\csname url@samestyle\endcsname
\providecommand{\newblock}{\relax}
\providecommand{\bibinfo}[2]{#2}
\providecommand{\BIBentrySTDinterwordspacing}{\spaceskip=0pt\relax}
\providecommand{\BIBentryALTinterwordstretchfactor}{4}
\providecommand{\BIBentryALTinterwordspacing}{\spaceskip=\fontdimen2\font plus
\BIBentryALTinterwordstretchfactor\fontdimen3\font minus
  \fontdimen4\font\relax}
\providecommand{\BIBforeignlanguage}[2]{{%
\expandafter\ifx\csname l@#1\endcsname\relax
\typeout{** WARNING: IEEEtran.bst: No hyphenation pattern has been}%
\typeout{** loaded for the language `#1'. Using the pattern for}%
\typeout{** the default language instead.}%
\else
\language=\csname l@#1\endcsname
\fi
#2}}
\providecommand{\BIBdecl}{\relax}
\BIBdecl

\bibitem{shen2012overview}
Z.~Shen, A.~Papasakellariou, J.~Montojo, D.~Gerstenberger, and F.~Xu,
  ``{Overview of 3GPP LTE-Advanced Carrier Aggregation for 4G Wireless
  Communications},'' \emph{IEEE Communications Magazine}, vol.~50, no.~2, pp.
  122--130, 2012.

\bibitem{yilmaz2019overview}
O.~N. Yilmaz, O.~Teyeb, and A.~Orsino, ``{Overview of LTE-NR dual
  connectivity},'' \emph{IEEE Communications Magazine}, vol.~57, no.~6, pp.
  138--144, 2019.

\bibitem{ghosh20195g}
A.~Ghosh, A.~Maeder, M.~Baker, and D.~Chandramouli, ``{5G evolution: A view on
  5G cellular technology beyond 3GPP release 15},'' \emph{IEEE access}, vol.~7,
  pp. 127\,639--127\,651, 2019.

\bibitem{lopez2019opportunities}
A.~V. Lopez, A.~Chervyakov, G.~Chance, S.~Verma, and Y.~Tang, ``{Opportunities
  and challenges of mmWave NR},'' \emph{IEEE Wireless Communications}, vol.~26,
  no.~2, pp. 4--6, 2019.

\bibitem{liu20205g}
G.~Liu, Y.~Huang, Z.~Chen, L.~Liu, Q.~Wang, and N.~Li, ``{5G deployment:
  Standalone vs. Non-standalone from the operator perspective},'' \emph{IEEE
  Communications Magazine}, vol.~58, no.~11, pp. 83--89, 2020.

\bibitem{gonzalez2017millimeter}
N.~Gonz{\'a}lez-Prelcic, A.~Ali, V.~Va, and R.~W. Heath, ``{Millimeter-wave
  communication with out-of-band information},'' \emph{IEEE Communications
  Magazine}, vol.~55, no.~12, pp. 140--146, 2017.

\bibitem{hashemi2017out}
M.~Hashemi, C.~E. Koksal, and N.~B. Shroff, ``{Out-of-band millimeter wave
  beamforming and communications to achieve low latency and high energy
  efficiency in 5G systems},'' \emph{IEEE transactions on communications},
  vol.~66, no.~2, pp. 875--888, 2017.

\bibitem{giordani2019standalone}
M.~Giordani, M.~Polese, A.~Roy, D.~Castor, and M.~Zorzi, ``{Standalone and
  non-standalone beam management for 3GPP NR at mmWaves},'' \emph{IEEE
  Communications Magazine}, vol.~57, no.~4, pp. 123--129, 2019.

\bibitem{yang2021integrated}
J.~Yang, X.~Yang, C.-K. Wen, and S.~Jin, ``Integrated sensing and communication
  with multi-domain cooperation,'' \emph{arXiv preprint arXiv:2105.03065},
  2021.

\bibitem{3GPP38901}
{3GPP Technical Report 38.901}, ``Study on channel model for frequencies from
  0.5 to 100 {GHz} ({R}elease 16),'' Dec. 2019.

\bibitem{heath2018foundations}
R.~W. Heath~Jr. and A.~Lozano, \emph{Foundations of {MIMO}
  Communication}.\hskip 1em plus 0.5em minus 0.4em\relax Cambridge University
  Press, 2018.

\bibitem{Rappaport2014-mmwbook}
T.~S. Rappaport, R.~W. {Heath Jr.}, R.~C. Daniels, and J.~N. Murdock,
  \emph{Millimeter Wave Wireless Communications}.\hskip 1em plus 0.5em minus
  0.4em\relax Pearson Education, 2014.

\bibitem{stocker1993neural}
K.~Stocker, B.~Gschwendtner, and F.~Landstorfer, ``Neural network approach to
  prediction of terrestrial wave propagation for mobile radio,'' in \emph{IEE
  Proceedings H (Microwaves, Antennas and Propagation)}, vol. 140, no.~4, 1993,
  pp. 315--320.

\bibitem{chang1997environment}
P.-R. Chang and W.-H. Yang, ``Environment-adaptation mobile radio propagation
  prediction using radial basis function neural networks,'' \emph{IEEE Trans.
  Veh. Techn.}, vol.~46, no.~1, pp. 155--160, 1997.

\bibitem{bai2018predicting}
L.~Bai, C.-X. Wang, J.~Huang, Q.~Xu, Y.~Yang, G.~Goussetis, J.~Sun, and
  W.~Zhang, ``Predicting wireless {mmWave} massive {MIMO} channel
  characteristics using machine learning algorithms,'' \emph{Wireless Commun.
  and Mobile Computing}, 2018.

\bibitem{huang2018big}
J.~Huang, C.-X. Wang, L.~Bai, J.~Sun, Y.~Yang, J.~Li, O.~Tirkkonen, and
  M.~Zhou, ``A big data enabled channel model for {5G} wireless communication
  systems,'' \emph{IEEE Trans. Big Data}, vol.~6, no.~2, pp. 211--222, 2018.

\bibitem{9122601}
X.~{Zhao}, F.~{Du}, S.~{Geng}, Z.~{Fu}, Z.~{Wang}, Y.~{Zhang}, Z.~{Zhou},
  L.~{Zhang}, and L.~{Yang}, ``Playback of {5G} and beyond measured {MIMO}
  channels by an {ANN}-based modeling and simulation framework,'' \emph{IEEE J.
  Sel. Areas Commun.}, vol.~38, no.~9, pp. 1945--1954, 2020.

\bibitem{XiaRanMez2020}
W.~Xia, S.~Rangan, M.~Mezzavilla, A.~Lozano, G.~Geraci, V.~Semkin, and
  G.~Loianno, ``Millimeter wave channel modeling via generative neural
  networks,'' in \emph{Proc. IEEE Globecom Workshops.}, 2020, pp. 1--6.

\bibitem{degli2007measurement}
V.~Degli-Esposti, F.~Fuschini, E.~M. Vitucci, and G.~Falciasecca, ``Measurement
  and modelling of scattering from buildings,'' \emph{IEEE Transactions on
  Antennas and Propagation}, vol.~55, no.~1, pp. 143--153, 2007.

\bibitem{schecklman2011terahertz}
S.~Schecklman, L.~M. Zurk, S.~Henry, and G.~P. Kniffin, ``Terahertz material
  detection from diffuse surface scattering,'' \emph{Journal of Applied
  Physics}, vol. 109, no.~9, p. 094902, 2011.

\bibitem{ju2019scattering}
S.~Ju, S.~H.~A. Shah, M.~A. Javed, J.~Li, G.~Palteru, J.~Robin, Y.~Xing,
  O.~Kanhere, and T.~S. Rappaport, ``Scattering mechanisms and modeling for
  terahertz wireless communications,'' in \emph{ICC 2019-2019 IEEE
  International Conference on Communications (ICC)}.\hskip 1em plus 0.5em minus
  0.4em\relax IEEE, 2019, pp. 1--7.

\bibitem{xing2021millimeter}
Y.~Xing and T.~S. Rappaport, ``Millimeter wave and terahertz urban microcell
  propagation measurements and models,'' \emph{IEEE Communications Letters},
  vol.~25, no.~12, pp. 3755--3759, 2021.

\bibitem{huang2020multi}
J.~Huang, C.-X. Wang, H.~Chang, J.~Sun, and X.~Gao, ``{Multi-frequency
  multi-scenario millimeter wave MIMO channel measurements and modeling for B5G
  wireless communication systems},'' \emph{IEEE Journal on Selected Areas in
  Communications}, vol.~38, no.~9, pp. 2010--2025, 2020.

\bibitem{cui2020multi}
Z.~Cui, C.~Briso-Rodriguez, K.~Guan, Z.~Zhong, and F.~Quitin,
  ``{Multi-frequency air-to-ground channel measurements and analysis for UAV
  communication systems},'' \emph{IEEE access}, vol.~8, pp. 110\,565--110\,574,
  2020.

\bibitem{khawaja2017uav}
W.~Khawaja, O.~Ozdemir, and I.~Guvenc, ``{UAV Air-to-Ground Channel
  Characterization for mmWave Systems},'' in \emph{Proc.\ IEEE VTC-Fall}, 2017.

\bibitem{Alkhateeb2019}
A.~Alkhateeb, ``{DeepMIMO}: A generic deep learning dataset for millimeter wave
  and massive {MIMO} applications,'' in \emph{Proc. of Information Theory and
  Applications Workshop (ITA)}, San Diego, CA, Feb 2019, pp. 1--8.

\bibitem{Remcom}
``{Remcom},'' available on-line at {\tt https://www.remcom.com/}, 2021-11-11.

\bibitem{he2018clustering}
R.~He, B.~Ai, A.~F. Molisch, G.~L. Stuber, Q.~Li, Z.~Zhong, and J.~Yu,
  ``Clustering enabled wireless channel modeling using big data algorithms,''
  \emph{IEEE Communications Magazine}, vol.~56, no.~5, pp. 177--183, 2018.

\bibitem{shafi20175g}
M.~Shafi, A.~F. Molisch, P.~J. Smith, T.~Haustein, P.~Zhu, P.~De~Silva,
  F.~Tufvesson, A.~Benjebbour, and G.~Wunder, ``{5G: A Tutorial Overview of
  Standards, Trials, Challenges, Deployment, and Practice},'' \emph{IEEE
  Journal on Selected Areas in Communications}, vol.~35, no.~6, pp. 1201--1221,
  2017.

\bibitem{giordani2020toward}
M.~Giordani, M.~Polese, M.~Mezzavilla, S.~Rangan, and M.~Zorzi, ``Toward 6g
  networks: Use cases and technologies,'' \emph{IEEE Communications Magazine},
  vol.~58, no.~3, pp. 55--61, 2020.

\bibitem{rappaport2019wireless}
T.~S. Rappaport, Y.~Xing, O.~Kanhere, S.~Ju, A.~Madanayake, S.~Mandal,
  A.~Alkhateeb, and G.~C. Trichopoulos, ``{Wireless communications and
  applications above 100 GHz: Opportunities and challenges for 6G and
  beyond},'' \emph{IEEE access}, vol.~7, pp. 78\,729--78\,757, 2019.

\bibitem{rousseeuw1987silhouettes}
P.~J. Rousseeuw, ``Silhouettes: a graphical aid to the interpretation and
  validation of cluster analysis,'' \emph{Journal of computational and applied
  mathematics}, vol.~20, pp. 53--65, 1987.

\bibitem{gulrajani2017improved}
I.~Gulrajani, F.~Ahmed, M.~Arjovsky, V.~Dumoulin, and A.~Courville, ``Improved
  training of wasserstein gans,'' \emph{arXiv preprint arXiv:1704.00028}, 2017.

\bibitem{arjovsky2017wasserstein}
M.~Arjovsky, S.~Chintala, and L.~Bottou, ``Wasserstein generative adversarial
  networks,'' in \emph{International conference on machine learning}.\hskip 1em
  plus 0.5em minus 0.4em\relax PMLR, 2017, pp. 214--223.

\bibitem{mmwchanmod-GAN}
``{mmWave} channel modeling git hub repository,'' available on-line at {\tt
  https://github.com/klmyyaqihu/mmwchanmod-GAN}, 2021-12-14.

\end{thebibliography}

\end{document}